\documentclass[12pt, preprint]{aastex}
\usepackage{natbib}
\newcommand{\ts}{t_{\rm stop}}
\newcommand{\vph}{v_{\rm wave}}
\newcommand{\Usum}{U_{\rm sum}}

\slugcomment{Draft Modified \today}

\shorttitle{Streaming Instabilities}
\shortauthors{Youdin \& Goodman}

\begin{document}

\title{Streaming Instabilities in Protoplanetary Disks}

\author{Andrew N. Youdin \& Jeremy Goodman}
\affil{Princeton University Observatory, Princeton, NJ 08544}

\begin{abstract}
Interpenetrating streams of solids and gas in a Keplerian disk produce a local, linear instability.  The two components mutually interact via aerodynamic drag, which generates radial drift and triggers unstable modes.  The secular instability does not require self-gravity, yet it generates growing particle density perturbations that could seed planetesimal formation.  Growth rates are slower than dynamical, but faster than radial drift, timescales.  Growth rates, like streaming velocities, are maximized for marginal coupling (stopping times comparable dynamical times).    Fastest growth occurs when the solid to gas density ratio is order unity and feedback is strongest.   Curiously, growth is strongly suppressed  when the densities are too nearly equal.  The relation between background drift and wave properties is explained by analogy with Howard's semicircle theorem.  The three-dimensional, two-fluid equations describe a sixth order (in the complex frequency) dispersion relation.  A terminal velocity approximation allows simplification to an approximate cubic dispersion relation.   To describe the simplest manifestation of this instability, we ignore complicating (but possibly relevant) factors like vertical stratification, dispersion of particle sizes, turbulence, and self-gravity.   We consider applications to planetesimal formation and compare our work to other studies of particle-gas dynamics.
\end{abstract}


\section{Introduction}
A major uncertainty in theories of planet formation occurs embarrassingly early, during the formation of planetesimals.  Collisions are unlikely to result in coagulation over a wide range of sizes, from mm to km, since available binding energies (chemical or gravitational) are negligible comparable to kinetic energies (\citealp{cth93, ys02}, hereafter YS; \citealp{you03}).  Zero gravity experiments confirm destructive cratering during low velocity impacts \citep{col03}.  The hypothesis that planetesimals form by gravitational collapse of solids that settle to the disk midplane (\citealp{gw73}; YS) remains controversial because it is uncertain whether protoplanetary disks are ever suitably quiescent or metal rich enough to allow gravitational instabilities to develop.

The strong coupling between small solids and gas is both an obstacle to, and a necessary ingredient in, planetesimal formation theory.  Forming planetesimals via gravitational instabilities would be trivial in a gas free disk, since collisional damping dominates viscous stirring in the absence of protoplanets \citep{gls04}.  Indeed \citet {gls04b} discuss a second generation of planetesimals that could form in a gas free environment during the final stages of planet formation.  However the first generation of planetesimals likely formed within the gas rich disks are observed to persist for several million years around low mass stars \citep{hll01}.  The outer planets of our solar system contain substantial amounts of this gas  (Uranus and Neptune have several earth mass atmospheres, while Jupiter and Saturn are mostly gas) which presumably accreted onto solid cores assembled from planetesimals that, by this logic, must have formed in the presence of gas.  While only trace amounts of gas are found on terrestrial planets, the simplest hypothesis is that inner-disk planetesimals also formed in the presence of gas.  Indeed conditions for gravitational instability become favorable when the gas disk is only partly dissipated (\citealp{sek98}, YS).

The relative abundances of solids and gas are not yet tightly constrained by observations.   At solar abundances the ratio of condensible solids to gas is of order $0.01$, and is higher or lower depending on whether ices condense.  The solid to gas surface density ratio, $\Sigma_p/\Sigma_g$, need not be fixed at solar abundances.  YS and \citet{yc04}, hereafter YC, discuss and model enrichment mechanisms (e.g.\ radial migration and photoevaporation) that act to increase $\Sigma_p/\Sigma_g$.  The ratio of space densities, $\rho_p/\rho_g$, is larger than $\Sigma_p/\Sigma_g$ toward the midplane of a stratified disk.  The extent of particle settling is limited by settling times (longer than disk lifetimes for sub-$\mu$m grains) and by turbulent diffusion, which can be generated by vertical stratification itself, among other possibilities.  Assuming that Kelvin-Helmoltz instabilities trigger this stirring, \citet{sek98} and YS showed that if $\Sigma_p/\Sigma_g$ is augmented by a factor $\sim 10$ the particle layer becomes so stratified that the gas and particle masses are equal in the particle sublayer, i.e. $\rho_p \sim \rho_g$ throughout the layer.  For even higher concentrations of solids, i.e.\ when the layer becomes particle-dominated, vertical shear instabilities are no longer capable of preventing particle settling and gravitational instabilities appear inevitable (YS).

Given that interesting effects occur when the solid and gas densities become comparable, Newton's Third Law tells us that we must consider the effects of frictional coupling on solids and gas equally.  However most studies of midplane particle dynamics (see \S\ref{comp}) do not fully treat the feedback of drag on the gas fluid.  We consider a simplified model  which treats the dynamics of both components self-consistently.  We investigate the linear stability properties of a two fluid Keplerian disk, where a pressureless fluid represents particles of a specific size.  It is well known that this system leads to steady state drift as angular momentum is transferred from the solids to the  pressure-supported, and thus sub-Keplerian, gas (\citealp{nsh86}, see \S\ref{basic}).  The radial component of this drift globally redistributes solids on long timescales, leading to ``particle pileups" since the accretion rate of solids is faster in outer disk (YS, YC).  

Here, however, we are concerned with local and more rapid consequences of orbital drift.  By analogy with the two stream instability in plasma physics \citep{spi65}, coupling  between interpenetrating streams destabilizes linear waves.  In our case the streams interact by drag forces, not electric fields.  Our model does not include self-gravity.  Nevertheless, unstable waves generate particle density perturbations.  In principle these perturbations could be relevant to planetesimal formation, for instance by raising the particle density to a point where self-gravity induces collapse of the perturbations.  We caution that the actual manifestation of particle-gas coupling may differ significantly from our model due to several simplifications.  We ignore vertical structure in our background state, so our system is effectively an infinite cylinder and not a thin disk.  Such an approximation may be warranted for vertical wavelengths smaller than disk scaleheights.  Furthermore our model is linear, laminar, and inviscid, though a possible non-linear outcome of the instability is weak turbulence.

This paper is organized as follows. Model equations and assumptions, for both steady state and perturbed motions, are presented in \S \ref{basic}.   Growth rates arising from the sixth order dispersion relation are numerically analyzed in \S\ref{results}.  The relation between growth rates and wave speeds are studied in \S\ref{sec:howard} by analogy with Howard's semicircle theorem.  Eigenfunctions of a vertically standing waves are constructed in \S\ref{motions}, allowing visualization of the fluid motions.  An approximate cubic dispersion that reproduces most features of the growing modes is derived in \S\ref{sec:cubic}, allowing analytic investigation to complement the results of \S\ref{results}.  Astrophysical applications considered in \S\ref{sec:app} include particle concentration (\S\ref{conc}) and a comparison of growth rates to steady state drift (\S\ref{sec:gvd}).  We compare our work to other studies of dust layer dynamics in \S\ref{comp}.  A summary and conclusions are given in \S\ref{disc}. 

\section{Basic Equations}\label{basic}

\begin{table}[tbp]
\caption{Symbols}
\begin{center}
\begin{tabular}{|c|c|c|}
\hline
Symbol & Definition & Meaning \\
\hline
$\ts$ & eqns.\ (\ref{Ep}, \ref{St}) & particle stopping time \\
$\tau_s$ & $\Omega \ts$ & dimensionless stopping time \\
$\rho_p$, $\rho_g$& & particle, gas space density\\
$\rho$ & $\rho_p + \rho_g$ & total density\\
$f_p$, $f_g$ & $\rho_p/\rho$, $\rho_g/\rho$ & particle, gas fraction \\
$\omega$ & & complex wave frequency \\
$s$, $\omega_\Re$ & $\Im(\omega)$, $\Re(\omega)$ & growth rate, wave frequency \\
$\vph$ & $\omega_\Re/k_x$ & radial wave (phase) speed\\
$k_x$, $k_z$ & &radial, vertical wavenumbers\\
$k$ & $\sqrt{k_x^2 + k_z^2}$ & wavenumber\\
$K_x$, $K_z$ & $k_x \eta r$, $k_z \eta r$ & dimensionless wavenumbers\\
$\eta$ & eqn.\ (\ref{eta}) & pressure parameter\\
$r$ & & cylindrical disk radius \\
$x$, $y$, $z$ & & rotating Cartesian grid \\
$V_K$ & $\Omega_K r$ & Keplerian circular speed \\
$\Omega$ & $(1- f_g\eta)\Omega_K$ & COM orbital frequency\\
${\bf V}_p$, ${\bf V}_g$ & & particle, gas fluid velocities\\
${\bf V}$ & $f_p {\bf V}_p+f_g {\bf V}_g$& COM velocity\\
$\Delta\!{\bf V}$ & ${\bf V}_p- {\bf V}_g$& relative velocity\\
${\bf v}$ & eqn.\ (\ref{Vpert}) & perturbed COM velocity\\
$\Delta{\bf v}$ & eqn.\ (\ref{DelVpert}) & perturbed relative velocity\\
$\delta$ & eqn.\ (\ref{rhopert}) & perturbed density\\
$h$ & eqn.\ (\ref{Ppert}) & perturbed pressure/enthalpy\\
\hline
\end{tabular}
\end{center}
\label{default} 
\end{table}

Our gas and particle ``fluids" obey continuity and Euler equations for the evolution of particle (${\bf V}_p$) and incompressible\footnote{Since motions are very subsonic, this assumption filters sound waves from the analysis.} gas (${\bf V}_g$) velocity, here presented in a non-rotating frame:
\begin{eqnarray}
{\partial \rho_p \over \partial t} + \nabla \cdot(\rho_p  {\bf V}_p) &=& 0\, , \label{contp}\\
\nabla \cdot {\bf V}_g &=& 0 \, , \\
{\partial {\bf V}_p \over \partial t} + {\bf V}_p\cdot\nabla {\bf V}_p &=& -\Omega_K^2 {\bf r} - {{\bf V}_p - {\bf V}_g \over \ts}\, , \label{Vp} \\
{\partial {\bf V}_g \over \partial t} + {\bf V}_g\cdot\nabla {\bf V}_g &=& -\Omega_K^2{\bf r} +{\rho_p \over \rho_g}{{\bf V}_p - {\bf V}_g \over \ts}-{\nabla P \over \rho_g} \, \label{Vg},
\end{eqnarray}
where $P$ is the gas pressure, $\rho_p$ and $\rho_g$ are the particle and gas spatial densities, respectively, and $\Omega_K \propto r^{-3/2}$ is the Keplerian orbital frequency at cylindrical radius $r$.  We ignore vertical stratification and self gravity for a simpler analysis, avoiding in particular the vertical settling and stirring of particles.  The particle stopping time, $\ts$, is conveniently independent of $\rho_p$, ${\bf V}_p$ and ${\bf V}_g$, for the small particles (radius $\ll 1$ m at 1 AU) of interest prior to planetesimal formation.  Epstein's law:
\begin{equation}\label{Ep}
\ts^{\rm Ep} = {\rho_s a \over \rho_g c_g}
\end{equation}
applies when $a < (4/9) \lambda_{\rm mfp}$,
where $a$ is the particle radius, $\lambda_{\rm mfp}$ is the gas mean free path (and $\lambda_{\rm mfp} \sim 1$ cm at $1$ AU), $c_g$ the gas sound speed, and $\rho_s$ denotes the material density of the solid.  Particles larger than $(4/9) \lambda_{\rm mfp}$, but small enough that the Reynolds number of the flow past the solid, $Re \equiv 4 a|{\bf V}_p - {\bf V}_g|/(c_g \lambda_{mfp}) < 1$ obey Stokes' law:
\begin{equation}\label{St}
\ts^{\rm St} = {4 \rho_s a^2 \over 9 \rho_g c_g \lambda_{mfp}}\, .
\end{equation}
For generality we use the dimensionless stopping time parameter:
\begin{equation}
\tau_s \equiv  \Omega_K \ts
\end{equation}
instead of referring to specific particle sizes and disk models. 
 
In this context,\footnote{A fluid description might also be possible given frequent interparticle collisions, but for small solids in a gas disk the stopping time is shorter than the collision time.}fluid description of particle motions (as opposed to the kinetic theory approach) requires that solids be tightly coupled to gas.  The criterion, $\tau_s \ll 1$, ensures strong coupling to dynamical perturbations, while $\omega \ts \ll 1$, suffices for disturbances of arbitrary frequency, $\omega$.  We will not consider $\tau_s > 1$.

Since relative motions between solids and gas are slow compared to center of mass velocities (in equilibrium and for perturbations), we express equations (\ref{Vp}, \ref{Vg}) in terms of relative motion, $\Delta\! {\bf V} \equiv {\bf V}_p - {\bf V}_g$, and center of mass (henceforth COM) motion, ${\bf V} \equiv (\rho_p {\bf V}_p + \rho_g{\bf V}_g)/\rho$, with $\rho = \rho_p + \rho_g$ the total density:
\begin{eqnarray}
{\partial {\bf V} \over \partial t} + {\bf V}\cdot\nabla {\bf V} + {\bf F}(\Delta\! {\bf V}^2) = - \Omega_K^2{\bf r} - \nabla P/\rho \, , \label{Vtot} \\
{\partial \Delta\!{\bf V} \over \partial t} + {\bf V}\cdot\nabla (\Delta\!{\bf V})+ \Delta\!{\bf V} \cdot\nabla {\bf V} + {\bf G}(\Delta\! {\bf V}^2) = -{\rho \over \rho_g}{\Delta\!{\bf V} \over \ts}+{\nabla P \over \rho_g} \, .\label{DelV}
\end{eqnarray}
The functions
\begin{eqnarray}\label{F}
{\bf F}(\Delta\! {\bf V}^2) &\equiv& {1 \over \rho} \nabla \cdot \left({\rho_g \rho_p \over \rho} \Delta\!{\bf V}\Delta\!{\bf V} \right) \, \\
{\bf G}(\Delta {\bf V}^2) &\equiv&  {\rho_g \over \rho} \Delta{\bf V} \cdot \nabla \left( {\rho_g \over \rho} \Delta {\bf V}\right)  - {\rho_p \over \rho}\Delta  {\bf V}\cdot\nabla \left( {\rho_p \over \rho}\Delta{\bf V}\right)\, , \label{G}
\end{eqnarray}
can often be dropped due to the smallness of ${\bf \Delta\!V}$, in which case equations  (\ref{Vtot}, \ref{DelV}) simplify considerably.  (As the conditions for the neglect of these terms differs for equilibrium and perturbed motions, they will be addressed subsequently.)  Another advantage of this formulation is that drag forces do not appear in COM evolution, and gravity, including self gravity were it included, is absent from the relative motion equation.  

\subsection{Steady Drift Solutions}\label{steady}
The time steady, axisymmetric drift motions of solids and gas have been well studied \citep{nsh86}, and are simple to rederive from equations (\ref{Vtot}, \ref{DelV}) with ${\bf F} = {\bf G} = 0$:
\begin{eqnarray}
\overline{U} &=& \overline{W} = \overline{\Delta W} = 0 \label{zeros}\\
\overline{V}  &=& \sqrt{V_K^2 + {r \over \rho}{\partial P \over \partial r}} \approx \left(1 - {\rho_g \over \rho} \eta\right)V_K  \, , \label{Vbar}\\
\overline{\Delta U} &=& -2 {\rho_g \over \rho}{\eta V_K\tau_s  \over 1 + (\tau_s \rho_g/\rho)^2}\, , \label{dVrbar}\\
\overline{\Delta\! V} &=& \left({\rho_g \over \rho}\right)^2{\eta V_K\tau_s^2 \over 1 + (\tau_s \rho_g/\rho)^2}\, ,\label{dVphibar}
\end{eqnarray}
where ${\bf V} = U\hat{r} + V\hat{\phi} + W\hat{z}$,  $\Delta\!{\bf V} = \Delta U\hat{r} + \Delta\! V\hat{\phi}+ \Delta W\hat{z}$, overbars denote steady state solutions, $V_K = \Omega_K r$ is the Keplerian circular speed, and
\begin{equation}\label{eta}
\eta \equiv- {1 \over 2\rho_g V_K^2}{\partial P \over \partial \ln r} \sim \left(c_g \over V_K\right)^2
\end{equation}
measures the radial pressure support.  In standard models of planet forming disks, $\eta \sim 10^{-3}$ at 1 AU.  Neglect of the non-linear drift terms (${\bf F}$, ${\bf G}$) is, in practice, always justified for equilibrium solutions.  These terms would only merit inclusion in the unlikely scenario that pressure strongly dominated gravity, $\eta \gg 1$.

Our equilibrium solutions (\ref{zeros} --- \ref{dVphibar}) contain no vertical motion, due to the neglect of vertical gravity.   The center of mass is fixed in radius, orbiting as a gas supported by a pressure $P$, but with a mean molucular weight augmented by $\rho/\rho_g$.    For outwardly decreasing pressure, $\eta > 0$, sub-Keplerian gas robs particles of angular momentum, giving inward migration of solids, $\overline{U}_p =  (\rho_g/\rho)\overline{\Delta U} < 0$, and outward migration of gas, $\overline{U}_g = - (\rho_p/\rho)\overline{\Delta U}$.  As it will set the scale for wave speeds, we introduce the unweighted sum of the radial drift speeds:
\begin{equation}\label{Usum}
\Usum \equiv \overline{U}_p + \overline{U}_g = -2 \rho_g {\rho_g -\rho_p \over \rho^2}{\eta V_K\tau_s \over 1 + (\tau_s \rho_g/\rho)^2}
\end{equation}
which has an interesting density dependence.  In the test particle limit, $\rho_p \rightarrow 0$, $\Usum \rightarrow \overline{U}_p \rightarrow -2\eta V_K \tau_s$.  With increasing particle concentration, the gas migrates faster at the expense of the solids, until $\Usum = 0$ for equal densities.  At $\rho_p = 3 \rho_g$, $\Usum = \eta V_K \tau_s/4$ reaches its (positive) maximum.  For even larger particle concentrations $\Usum$ declines as the gas pressure is diluted by the particle mass.
 
Azimuthal drift is much slower than radial, $\overline{\Delta V}/\overline{\Delta U} = \rho_g \tau_s/(2 \rho)$ for tight coupling.  For loose coupling,  $\overline{\Delta V} \rightarrow \eta V_K$ as particle and gas trajectories approach unperturbed Keplerian and pressure supported rotation, respectively.  The previous caution about applying fluid equations for $\tau_s \gg 1$ can be ignored for these equilibrium solutions provided the disk varies negligibly over a stopping length $\ts \overline{\Delta U} \sim \eta r \ll r$.  Finally note that our unstratified model, with $\partial \overline{V}/\partial z = 0$, ignores the vertical shear, which can generate Kelvin-Helmholtz instabilities.

\subsection{Localized Perturbation Equations}
We consider perturbations to steady state motion (equations \ref{Vbar} --- \ref{dVphibar}) on length scales much shorter than disk radii (indeed, shorter than the thickness of the gas layer, $H_g \sim \eta^{1/2}r$).  This allows a local treatment, with Cartesian coordinates corotating about a fixed radius, $r_o$, with the orbital frequency of the local COM,  $\Omega_o  = \overline{V}_{\phi}(r_o)/r_o$.  In the new coordinates:
\begin{eqnarray}
x \equiv r - r_o \, , \\
y \equiv r_o(\phi - \Omega_o t)  \, ,
\end{eqnarray}
we approximate differential rotation, as usual, by plane parallel flow with linear shear,  $\overline{\bf V} = -q\Omega_o x \hat{y}$, where $q \simeq 3/2$ for the nearly Keplerian profile.  Within this approximation drift motions are radially constant,  $d\overline{\Delta\!{\bf V}}/dx = 0$.  We decompose fluid variables into steady backgrounds and perturbations:
\begin{eqnarray}
{\bf V} &=& -\textrm{$\frac{3}{2}$}\Omega_o x \hat{y} + {\bf v}(x,z,t)\, , \label{Vpert} \\
 \Delta\! {\bf V} &=&  \overline{\Delta U}\hat{x} + \overline{\Delta V}\hat{y} + \Delta\! {\bf v}(x,z,t)\, , \label{DelVpert}\\
\rho &=& \rho_{o}[1 + \delta(x,z,t)] \, ,\label{rhopert}\\
P &=& \rho_{o}[ -g_e x + h(x,z,t)] \, , \label{Ppert}
\end{eqnarray}
where $g_e =  -dP_o/dr|_{r_o}/\rho_{o} = 2\eta\Omega_o^2r_o\rho_g/\rho_o>0$.  Henceforth we drop overbars and the subscripted $o$'s from unperturbed states.  Perturbations are axisymmetric, which avoids stretching by radial shear, and are given a Fourier dependence:
\begin{equation}\label{fourier}
f(x,z,t) = \tilde{f} \exp[\imath(k_x x + k_z z - \omega t)] \, ,
\end{equation}
with real wavenumbers, $k_x$ and $k_z$, and a complex frequency,
\begin{equation}
\omega = \omega_\Re + \imath s\, ,
\end{equation}
with a wave frequency, $\omega_\Re$, and  growth (or damping) rate, $s$.\footnote{Modes with $s>0$ and $\omega_\Re \ne 0$ are often called ``overstable" to distinguish them from non-oscillatory instabilities.}

The linear perturbation equations read:
\begin{eqnarray}
-\imath \omega {\bf v} -2\Omega v \hat{x} + {\Omega \over 2} u \hat{y} + {\bf F}' &=&
-\imath {\bf k}h - \delta g_e\hat{x} \label{v} \\
-\imath \omega \Delta\! {\bf v} -2\Omega \Delta v \hat{x} + {\Omega \over 2} \Delta u \hat{y}  + i k_x \Delta U {\bf v}  + {\bf G}' &=& - {\left(\Delta\! {\bf v}  +  \delta {\Delta\! {\bf V}}\right) \over f_g \ts}  +
\imath {\bf k} {h \over f_g} \label{dv}\\
-\imath \omega \delta +\imath {\bf k}\cdot {\bf v} &=& 0 \label{cont} \\
{\bf k}\cdot{\bf v} &=& f_p {\bf k}\cdot\Delta\! {\bf v} + {f_g  }k_x \Delta U \delta \label{incomp}
\end{eqnarray}
and the terms that arise from perturbations of equations (\ref{F}, \ref{G}) are:
\begin{eqnarray}
{\bf F}' &=& \imath f_g [(f_p {\bf k}\cdot\Delta\!{\bf v} + f_g k_x \Delta U \delta)\Delta\!{\bf V} + f_p k_x \Delta U \Delta\!{\bf v}] \label{F'}\\
{\bf G}' &=& \imath k_x \Delta U \left[(f_g - f_p)\Delta\! {\bf v} - f_g \Delta\! {\bf V}\delta \right]\label{G'}
\end{eqnarray}
where ${\bf v} = u\hat{x} + v\hat{y} + w\hat{z}$, $\Delta\!{\bf v} = \Delta u\hat{x} + \Delta v\hat{y} + \Delta w\hat{z}$,
$f_g \equiv \rho_g/\rho$ is the equilibrium gas fraction, and the particle fraction, $f_p = 1-f_g$.  

Equations (\ref{v} --- \ref{G'}) define a 6th order (one less than the number of time derivatives because of the incompressibility constraint) dispersion relation for $\omega$, whose solutions we investigate in \S\ref{results}.  Equivalent results are obtainted by perturbing  particle and gas equations (\ref{contp}---\ref{Vg}) directly, but the relative velocity formulation allows analytic simplifications.  For low-frequency waves with $\omega \lesssim \Omega$, it is safe to neglect ${\bf F}'$ and ${\bf G}'$, but all terms are included in our numerical solutions.  In \S\ref{sec:cubic}, further approximations allow us to derive an approximate cubic dispersion relation that describes the growing modes while filtering three strongly damped modes.
 
\section{Growth Rates and Wave Speeds}\label{results}
We investigate the growth rates and wave speeds, $\vph \equiv \omega_\Re/k_x$ of linear disturbances.  The eigenvalue problem for $\omega/\Omega$ (prescribed by equations \ref{v} --- \ref{incomp}) is uniquely specified by  four quantities:  the stopping time parameter, $\tau_s = \Omega \ts$, and the equilibrium solid-gas density ratio, $\rho_p/\rho_g$, which define the background state, and two normalized wavenumbers,  $K_x = \eta r k_x$ and $K_z = \eta r k_z$.  Of the six modes, three decay within a stopping time, and are of little physical interest.  The other three modes, of which two are modified epicycles and the other is a uniquely two fluid secular mode, can be slowly  damped or growing.  The secular mode gives the fastest growth.  This section investigates the fastest growing waves, considers whether turbulent viscosity can stabilize short wavelength modes, and relates growth rates and wave speeds to the background flow, by analogy with Howard's (1961) semicircle theorem.

\begin{figure}[tb]
\begin{center}
\centerline {
\includegraphics[width=7in]
{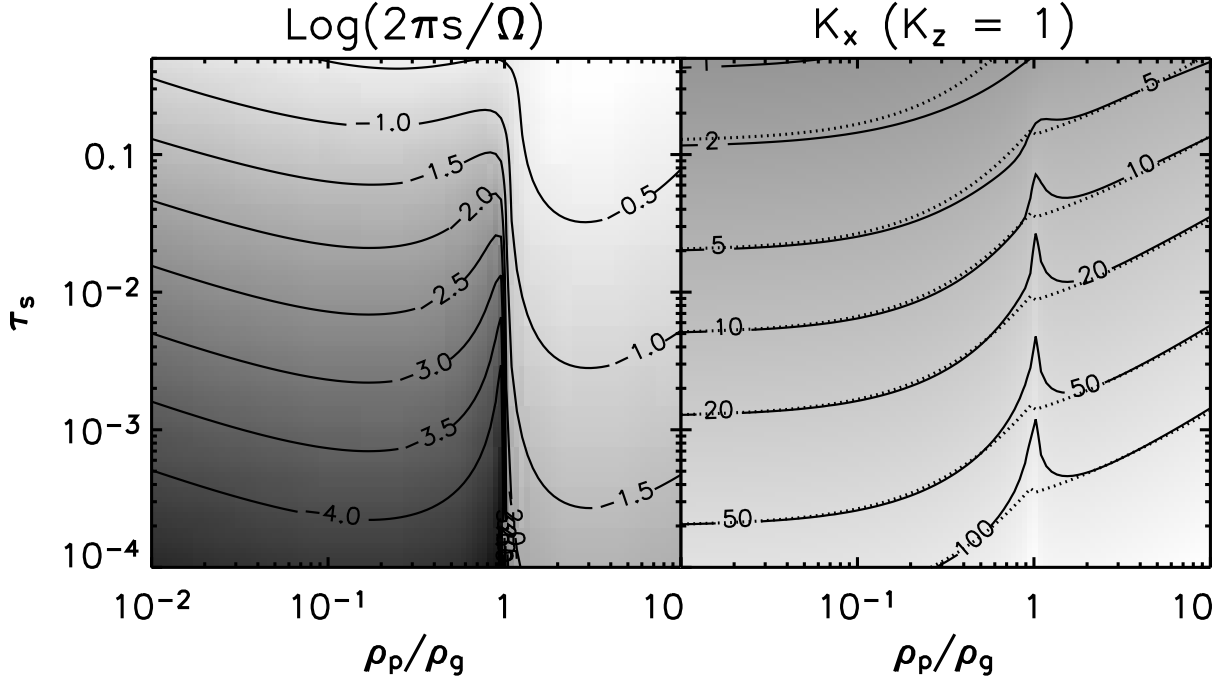}
}
\caption{(\emph{Left}) Contours of growth rate (times orbital period) vs.\ stopping time and solid to gas density ratio.  The vertical wavenumber is fixed at $K_z = 1$ while $K_x$ is varied to maximize growth.  (\emph{Right}) Radial wavenumbers (\emph{solid contours}) of these fastest growing modes.  The power law fits (\emph{dotted contours})  follow equation (\ref{powerlaw}).  See text (\S\ref{results}) for discussion.}
\label{maxgrowth}
\end{center}
\end{figure}


\subsection{General Features}
Figure \ref{maxgrowth} (\emph{left}) plots growth rate contours versus stopping time and $\rho_p/\rho_g$.  We hold $K_z = 1$ fixed (more on this choice later) and maximize the growth rate with respect to $K_x$, which is plotted at right.  Growth is possible for all values of $\tau_s$ and $\rho_p/\rho_g$, on slower than dynamical timescales.  In the well coupled regime, $\tau_s \ll 1$, peak growth rates increase as $s \propto \tau_s$ (for fixed $\rho_p/\rho_g$) since looser coupling increases the relative motion needed for instability.  The growth rates peak for marginal coupling and decrease for  $\tau_s \gtrsim 1$, but we ignore this regime where a fluid description of the solids is questionable.

The density dependence in figure \ref{maxgrowth} is more complicated, and has nothing to do with self-gravity, which is not included.  As expected, growth rates  decrease in the test-particle limit, as $\rho_p/\rho_g \rightarrow 0$, and as $\rho_p/\rho_g \rightarrow \infty$ when solids are unaffected by drag.  More surprisingly, growth is diminished in a narrow region around $\rho_{p} = \rho_g$.   We will show that this is related to vanishing wave speeds. Consequently, two lobes of relatively fast growth (around $\rho_p/\rho_g \approx 0.2$ and $3$) exist where particle gas feedback is significant, but not so close to $\rho_p = \rho_g$ that waves stagnate.  The fastest growth occurs in the particle-dominated lobe.  

The $K_x$ values of figure \ref{maxgrowth} (\emph{right}) are well fit by a broken power law:
\begin{equation}\label{powerlaw}
K_x = \left\{ \begin{array}{ll} 
(2 \tau_s f_g^3)^{-1/2} & \textrm{if $f_g < 1/2$}\\
\sqrt{2} \tau_s^{-1/2} f_g^{-0.4} & \textrm{if $f_g > 1/2$}
\end{array}
\right.
\end{equation}
except for the spike at $f_g = 1/2$.  The preferred radial wavelengths decrease with the \emph{particle} stopping time as $K_x \propto \tau_s^{-1/2}$.  The increase in $K_x$ with $\rho_p/\rho_g$ is related to the density dependence of the \emph{gas} stopping time, $\ts \rho_g/\rho_p$ (see equation \ref{Vg}), and stopping time for relative motions, $\ts \rho_g/\rho$ (see equation \ref{DelV}).   The two fluids become more tightly coupled as particle concentration increases, resulting in shorter wavelength growing modes.   

\subsection{The Long and Short of It}\label{kdep}

\begin{figure}[tb]
\begin{center}
\centerline{
\includegraphics[width=7in]
{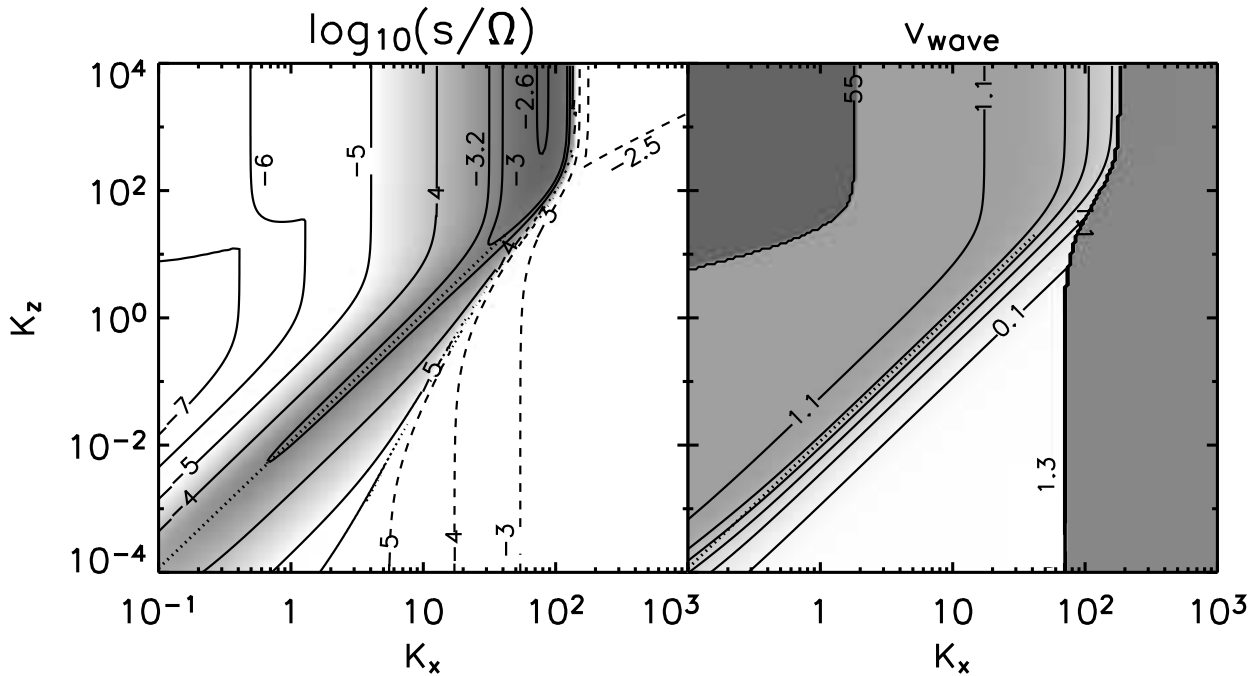}
}
\caption{(\emph{Left}) Contours of growth rate, $\log_{10}(s/\Omega)$ (\emph{solid lines}), and damping rate, $\log_{10}(-s/\Omega)$ (\emph{dashed lines}), vs. $K_x$ and $K_z$ for $\rho_p/\rho_g = 0.2$ and $\tau_s = 0.01$.  Two regions contain the fastest growing modes:  $K_z \gtrsim 10^2$, $K_x \approx 80$ (the darkly shaded region) and along $K_z \approx \tau_s K_x^2 f_g^3$ (the \emph{dotted line} in both figures).  (\emph{Right}) Wave speed, $\vph = \omega_\Re/k_x$, in units of $-\eta v_K \tau_s$, for the same modes.  The contours from $0.1$ to $1.1$ increment by $0.2$ and $s$ peaks along this gradient in phase speed.  Darkly shaded, large phase speed regions to the right (and upper left) correspond to damped (and very weakly growing) modes, respectively.}
\label{kxkz}
\end{center}
\end{figure}

Figure \ref{kxkz} (\emph{left}) plots growth rate versus wavenumbers ($K_x, K_z$).  Specific values of the density ratio and stopping time ($\rho_p/\rho_g = 0.2$, $\tau_s = 0.01$) are chosen, but the qualitative features are rather general.  We can identify two ``ridges" along which growth rates peak.  The short wavelength branch follows vertical contours ($K_x \approx 80$ and $K_z \gtrsim 100$ in the figure) while the long wavelength ridge falls diagonally along $K_z \sim \tau_s K_x^2$ (a generalization of equation \ref{powerlaw} for $K_z \ne 1$ that ignores the density dependence).  A smooth transition between the two ridges occurs around $K_x \sim K_z \sim 1/\tau_s$.  Very long wavelength modes ($K_x \lesssim 1$, $K_z \lesssim \tau_s$) are damped by frictional dissipation and angular momentum gradients.  Much of this paper chooses the long wavelength branch by fixing $K_z = 1$.  The exact value chosen is arbitrary as growth rates vary little along this ridge. This subsection analyzes the short waves which have larger growth rates, but could be less relevant if turbulent viscosity were included in the analysis.

\subsubsection{Short Wavelength Limit: $K_z \gg K_x$}

\begin{figure}[htb]
\begin{center}
\centerline{
\includegraphics[width=4in]
{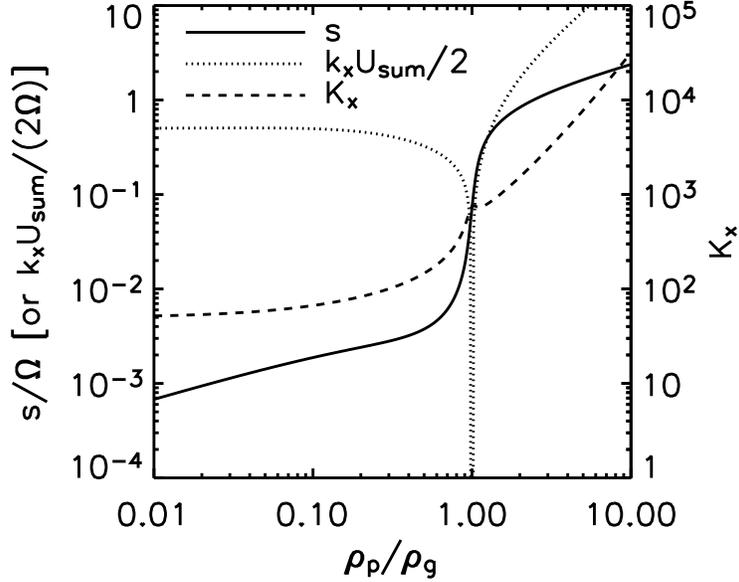}
}
\caption{ Maximum growth rates ({\it solid line}) and fastest growing radial wavenumber (\emph{dashed line}) versus solid to gas density ratio for $\tau_s = 0.01$ in the limit $K_z \gg K_x$.  The growth rates are below the upper limit implied by the semicircle theorem (\emph{dotted lines}), except for a narrow region near equal densities.}
\label{KzInf}
\end{center}
\end{figure}

The short wavelength behavior is described by a dispersion relation that is independent of $K_z$  for $K_z \gg K_x$, as seen in figure \ref{kxkz} .  Physically, radial pressure perturbations become negligible.  Figure \ref{KzInf} plots the maximum growth rates, and the $K_x$ values of the fastest growing modes, in this large $K_z$ limit.  For a gas-dominated system, the short waves grow marginally faster than the long wave branch, the difference is less than a factor of 5 in the gas dominated regime, $\rho_p/\rho_g \lesssim 0.2$.  The preferred radial wavenumber is nearly constant, $K_x \sim 1/\tau_s$, in the gas-dominated regime.  The growth rate skyrockets as the density ratio, $\rho_p /\rho_g$, increases toward and above unity.  This contrasts with the behavior we saw in the long wavelength case where growth is suppressed near equal densities.  The $K_x$ value that maximizes growth increases with particle fraction, and does not keep the characteristic value $1/\tau_s$.

The real frequencies of these modes (in both the particle- and gas-dominated regimes, not plotted) are near, but slightly below, the dynamical frequency.  Longer wavelength modes have lower frequencies.  Indeed, figure \ref{kxkz} shows that $\vph \equiv \omega_\Re/k_x$ is similar for the fastest modes at all wavelengths.  Hence modes with higher $k_x$ must have higher frequencies.  The terms in equation (\ref{F'}, \ref{G'}) must be kept to describe the short wavelength, high frequency modes.

\subsubsection{Turbulent diffusion}
Another factor to consider in determining the relative significance of short or long waves is turbulent diffusion.  Turbulence is not included in our model, and so we cannot be certain of its effects.  If it introduces viscous diffusion, short wavelength modes would be preferentially damped.  To estimate the relevance of this effect, consider the diffusive timescale, $t_D \sim 4\pi^2/(k^2 D)$ with the diffusion coefficient parametrized by the usual prescription $D \equiv \alpha c_g^2/\Omega$.  Growth outpaces diffusion if $s t_D \gtrsim 1$, or equivalently:
\begin{equation}\label{kmax}
K \lesssim 2 \pi \sqrt{s \eta \over \Omega \alpha}
\end{equation}
Considerable uncertainly surrounds the appropriate value for $\alpha$.  The values invoked to explain accretion onto young stars, ranging from $10^{-4}$ --- $10^{-2}$ at least, may not apply here.  Even if accretion is driven by turbulent diffusion, disks likely contain spatial inhomegeneities (disk midplanes which may be more quiescent) and experience temporal evolution (accretion rates decrease with age).

It is more relevant to consider diffusivities needed to maintain a thin, but finite density, particle layer.  A simple balance between settling and diffusion (see e.g.\ YC) for well-coupled particles ($\tau_s \ll 1$) suggests that $\alpha \sim \eta \tau_s[H_p/(\eta r)]^2$ is required for sedimentation to a particle scaleheight $H_p$ (which is assumed to be thinner than the gas scale height).  If $H_p \approx \eta r$ (thinner layers may be strongly Kelvin-Helmoltz unstable), equation (\ref{kmax}) gives $K \lesssim 2 \pi \sqrt{s/(\Omega \tau_s)}$.  Since $s \lesssim \Omega \tau_s$ (for $\rho_p < \rho_g$ at least), short modes with $K \gg 1$ should be strongly damped.  Even longer wavelength modes, with $K \sim 1$ --- $10$, are affected viscous diffusion, according to this analysis.  However, viscous effects can sometimes destabilize disks \citep{st95}. This issue merits further study.

\begin{figure}[htb]
\begin{center}
\centerline {
\includegraphics[width=7in]
{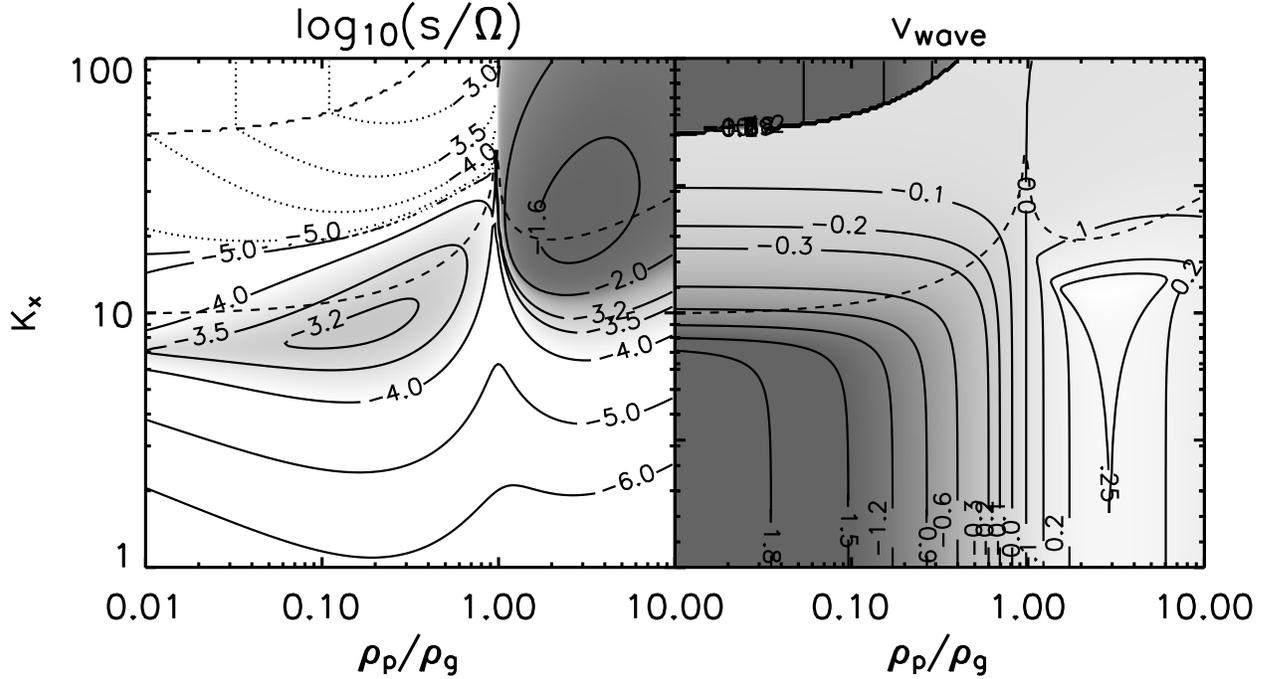}
}
\caption{(\emph{Left}) Growth rate, $\log_{10}(s/\Omega)$, (\emph{solid contours}) and decay rates, $\log_{10}(-s/\Omega)$  (\emph{dotted contours}) vs. solid to gas density ratio and radial wavenumber for $\tau_s = 0.01$ and $K_z = 1$.   Two lobes of rapid growth are centered on $\rho_p/\rho_g \approx 0.2$ and $3$, with suppressed growth near $\rho_p = \rho_g$.  (\emph{Right}) Radial wave speed contours in units of $\eta v_K \tau_s$ for the same modes.  The phase speed changes sign across $\rho_p = \rho_g$.  The \emph{dashed contours} in both plots indicate the location where $\vph = \Usum/2$.}
\label{mukx}
\end{center}
\end{figure}

\subsection{Phase Speeds and the Semicircle Theorem}\label{sec:howard}

 Figure \ref{kxkz} (\emph{bottom}) plots the wave speed, $\vph$, of the modes whose growth rates were shown in figure \ref{kxkz}.  For the mode of interest, we see two plateaus of nearly constant phase speed.\footnote{ We can ignore the discontinuities in the upper left corner and far right hand side of the plot, which correspond to different, epicyclic, roots that happen to give larger growth rates in this region of phase space, which is generally uninteresting as growth rates are small.}  The steep transition between these values overlaps the ridge of large growth rates in figure \ref{kxkz}.   

These results are analogous to Howard's semicircle theorem for the Kelvin-Helmholtz instability, which we summarize briefly (see \citet{kundu} for a derivation).  \citet{how61} found that one dimensional modes, $\propto \exp (s t - \imath \omega_\Re + \imath k_x x)$, have a wave speed that lies between the minimum and maximum speeds in the shearing flow, $V_{\rm min} < \omega_\Re/k < V_{\rm max}$.  The semicircle theorem:
\begin{equation}
 \left[\omega_\Re/k - \textrm{$\frac{1}{2}$} (V_{\rm max} + V_{\rm min})^2\right]^2 + (s /k)^2 \leq  \left[ \textrm{$\frac{1}{2}$}(V_{\rm max} - V_{\rm min})^2\right]^2 
\end{equation}
 says that the complex wave velocity of an unstable mode lies in a semicircle (since only positive growth rates are considered) of radius $V_{\rm max} - V_{\rm min}$.  This imposes a limit on the maximum growth rate:
 \begin{equation}
 s \leq  \textrm{$\frac{k}{2}$}(V_{\rm max} - V_{\rm min}),
 \end{equation}
which is achieved for a phase speed midway between the allowed range.

The physical differences between our streaming instability and the Kelvin-Helmholtz instability cannot be overstated:  two interpenetrating, unstratified, rotating fluids with 3D velocities and 2D wavenumbers versus a single, plane-parallel, neutrally buoyant fluid with 2D velocities and 1D wavenumbers.  It is remarkable then that our modes behave as if a modified semicircle theorem applied to them.   As no proof of an analogous theorem for our problem exists,  we describe the similarities.  First, the radial phase speeds of our secular growing mode fall in the range, $0 < | \vph| < |\Usum|$,\footnote{For $\rho_p > \rho_g$, the behavior is a bit more complicated.  This case will be discussed shortly.}
where $\Usum$ is the sum of particle and gas drift velocities, see equation (\ref{Usum}).  As $\Usum$ is positive (negative) in a particle (gas) dominated layer, respectively $\vph$ has the same sign as $\Usum$.  
Secondly, growth rates are largest near $\vph \approx \Usum/2$ and vanish near the endpoints of the allowed range.  Since $\Usum = 0$ for equal densities ($\rho_p = \rho_g$), the semicircle theorem is consistent with the finding that growth is weakened in this case (though not in the large wavenumber limit, as we have discussed).  The peak growth rates are indeed bounded by $s < |k_x \Usum|/2$ except for a very narrow region around $\rho_p = \rho_g$.

To demonstrate the generality of these findings, figure \ref{mukx} plots the growth rates and wave speeds, versus $K_x$ and $\rho_p/\rho_g$.  
At small $K_x$, wave speeds,  $\vph$, approach $\Usum$, for instance with $\rho_p/\rho_g = 0.1$, $\vph \rightarrow \Usum \simeq -1.5 \eta v_K \tau_s$ as the contour indicates.    For large $K_x$, $\vph \rightarrow 0$ 
 (ignoring the mode switching of the dark region in the upper left corner).  Growth rates are largest roughly midway through this transition, where $\vph \approx \Usum/2$, as indicated by the dashed contours.  The suppression of growth for $\rho_p = \rho_g$, when $\vph \simeq \Usum = 0$ and the ``radius" of the semicircle vanishes, is clear.

The transition in $\vph$ from $\Usum$ to $0$  with increasing $K_x$ has an added wrinkle in the particle-dominated case.  The wave speed first rises slightly above $\Usum$ (which is clearly not a strict upper limit) before dropping to zero, as can be seen by following the contours in figure (\ref{mukx}, \emph{right}) for $\rho_p/\rho_g > 1$.  The analogy to the semicircle theorem is still relevant, as the fastest growth occurs for $\vph \approx \Usum/2$.

The analogy to the semicircle theorem connects the background flow to wave properties.  The free energy of interpenetrating streams undoubtedly plays a role, but from this perspective, it is surprising that the velocity scale is set by the sum rather the difference of radial streaming velocities.  More study of two coupled, rotating fluids should increase our understanding of this system.  

\begin{figure}[tb!]
\begin{center}
\centerline {
\includegraphics[width=4in]
{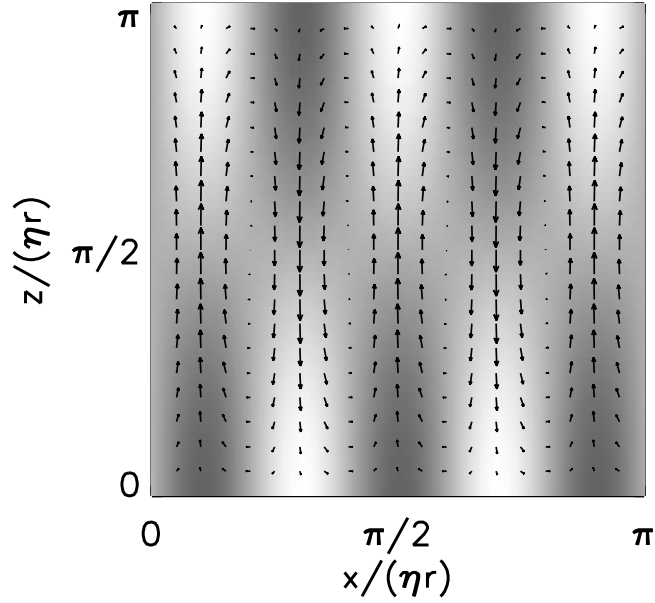}
}
\caption{Instantaneous (perturbed) particle velocity, ${\bf v}_p$, in the $x-z$ plane with a greyscale image of azimuthal velocities (white is positive) for a growing mode with  $K_x = 5$, $|K_z| = 1$, $\tau_s \approx .044$, $\rho_p/\rho_g = 0.2$.  Gas velocities are very similar due to strong coupling.  The density is very nearly in phase with the azimuthal speed, so the vertical flow is channeled to high density regions.  The ratio of azimuthal to vertical 
velocity amplitudes is $|v_p|/|w_p| \simeq 0.66$.  The radial to vertical ratio, $|u_p|/|w_p| \simeq K_z/K_x = 0.2$, follows from near incompressibility.  This mode has a growth rate $s/\Omega \approx 2.9 \times 10^{-3}$ and a phase speed, $\omega_\Re/k_x = -0.42 |\Delta U|$.}
\label{efunct}
\end{center}
\end{figure}
\begin{figure}[htb]
\begin{center}
\centerline {
\includegraphics[width=4in]
{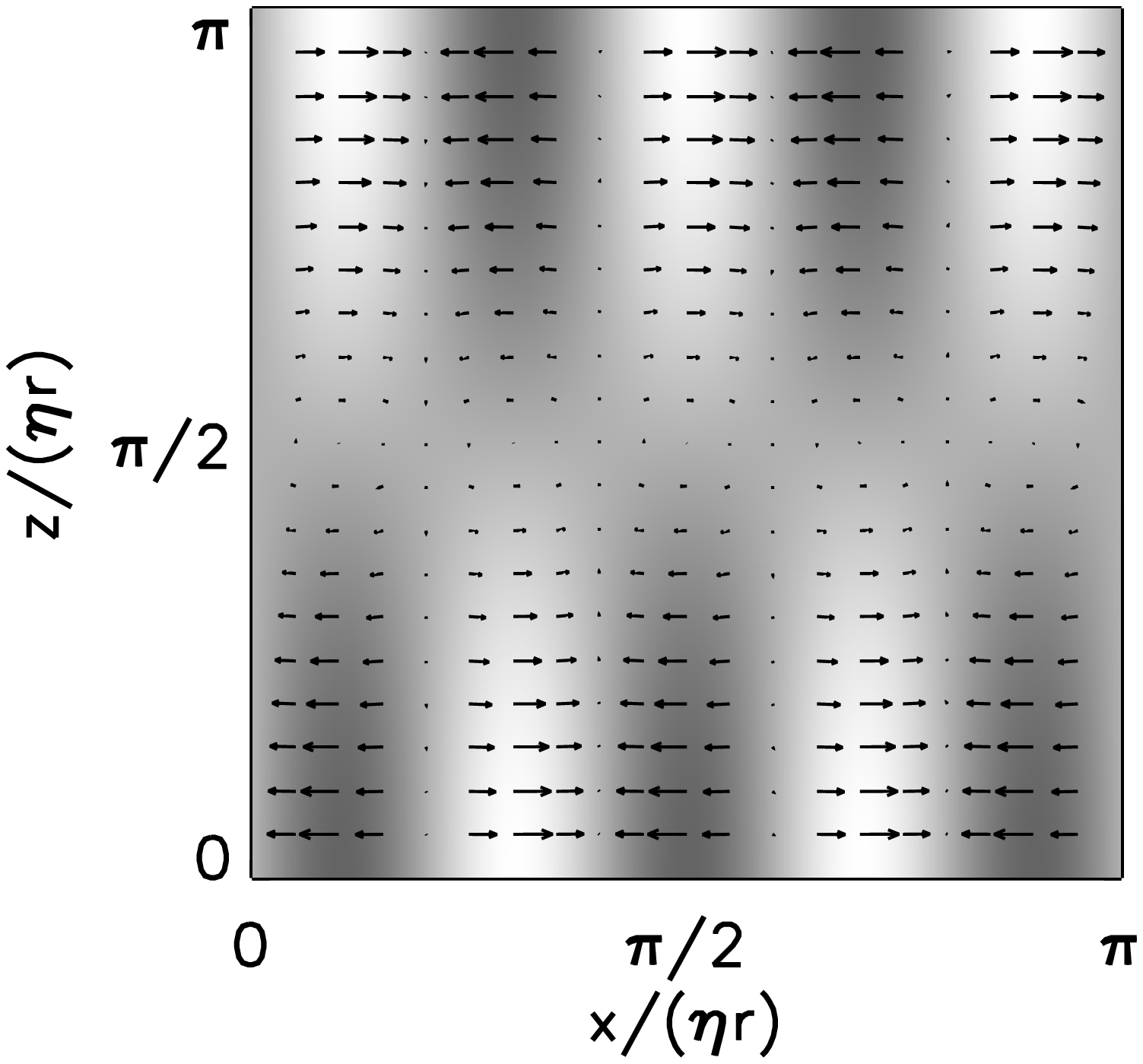}
}
\caption{Perturbed relative motion of solids and gas, $\Delta {\bf v}$, for the same mode as figure \ref{efunct}.  The greyscale image shows density perturbations (white is positive).  The radial relative motion dominates the azimuthal, $|\Delta v|/|\Delta u| \approx 0.15$, and vertical,  $|\Delta w|/|\Delta u| \approx 0.11$, speeds.  Density perturbations correlate with relative motion.}
\label{dvkx5}
\end{center}
\end{figure}

\section{Eigenfunctions: Fluid Motions and Density Perturbations}\label{motions}

Having investigated the eigenvalues, i.e.\ the growth rates and phase speeds of a mode, we consider the eigenfunctions, i.e. the Fourier amplitudes, $\tilde{f}$, that give us the fluid variables via equation (\ref{fourier}).  An individual mode has a vertical phase speed, $\omega_\Re/k_z$, which can be eliminated by linearly superposing pairs of modes with opposite signs of $k_z$.  Under a vertical parity transformation, the vertical velocity is odd, while $\omega$ and all other Fourier amplitudes are even.  The vertical standing waves have forms:
\begin{eqnarray}
f_{\rm odd} = \Re(\imath \tilde{f} \exp[\imath(k_x x  - \omega t)])\sin( k_z z)\, \label{odd} \\
f_{\rm even} = \Re(\tilde{f} \exp[\imath(k_x x  - \omega t)])\cos( k_z z)\, \label{even} 
\end{eqnarray}
for the odd (vertical velocity) and even (all other) variables, respectively.

Figure \ref{efunct} plots particle velocities for a rapidly growing mode.  The $K_x$ value gives the fastest growth rate for $K_z = 1$ as in figure \ref{maxgrowth}.   Since vertical wavelengths are longer than radial, tight coupling of particles to the incompressible gas, $\partial u_g/\partial x +\partial w_g/\partial z = 0$, causes vertical velocities to dominate.  Recall from figure \ref{maxgrowth} (\emph{right}) that elongation of the fastest growing modes is more (less) pronounced for tighter (looser) coupling.   The vertical velocities flow toward density maxima for this growing secular mode.  The pair of epicycles are weakly damped ($s \approx -2.4\times 10^{-3} \Omega$ and $s \approx -8.9\times 10^{-3} \Omega$)  for these parameters.  Their flow patterns are similar to figure \ref{efunct} except vertical velocities flow toward density minima.  Gas and particle velocities are not well coupled for the three strongly damped modes.  The gas is nearly stationary so particle motion leads to rapid decay.

Figure \ref{dvkx5} shows perturbed relative velocities for the same mode as figure \ref{efunct} with the perturbed density in greyscale.  This relative motion is predominantly radial, even accounting for azimuthal velocities.  The correlation of density with $\Delta u$ can be derived from continuity equation using $\omega_\Re >> s$.    

These eigenfunctions cannot be fit into a finite thickness dust layer.  This is clear from equations (\ref{odd} and \ref{even}) and figure \ref{efunct}  which show that vertical velocities are maximized where density (and other component of velocity) vanish, and vice versa.  A more complicated model that includes either stratification or a free surface between the particle layer and overlying gas layers, would give eigenfunctions with more realistic boundary conditions.

\section{Terminal Velocity Approximation}\label{sec:cubic}
A simpler description of unstable modes, which filters the three strongly damped roots, is achieved by assuming that relative velocities reach ``terminal velocity," so that drag forces adjust quasistatically to pressure forces.  This amounts to neglecting all terms on the left-hand side of equation (\ref{DelV}), both in equilibrium and in perturbation.  Thus $\Delta\!{\bf V} = -(\nabla P/\rho)\ts$, and perturbations obey
\begin{equation}
\Delta {\bf v}  +  \delta \Delta {\bf V}  \approx
\imath {\bf k} h \ts \label{dvsimp}\, .
\end{equation}
This approximation, which ignores inertial accelerations, is valid so long as $K  \ll 1/\tau_s$ and $\tau_s \ll 1$. 

A cubic dispersion relation:
\begin{equation}\label{cubic}
\left({\omega \over \Omega}\right)^3 + \tau_s \left({\omega \over \Omega}\right)^2 \left(\imath f_p{K_x^2 \over K^2}+ 2 f_g^2 K_x\right) -  \left({\omega \over \Omega}\right) \left({K_z \over K}\right)^2 + 2 f_g K_x\left({K_z \over K}\right)^2(f_p - f_g)\tau_s = 0
\end{equation}
results from equations (\ref{v}, \ref{cont}, \ref{incomp}, and \ref{dvsimp}), see the appendix for intermediate equations.
The roots of this cubic reproduce the results of the full system of equations to very good approximation when the stated assumptions are met.

\subsection{Stability of Inplane Motions}
When $K_z = 0$, equation (\ref{cubic}) gives:
\begin{equation}
 -\imath \omega/\Omega = - f_p  \tau_s - 2 \imath f_g^2 K_x \tau_s
\end{equation}
so that all modes are damped, $s < 0$.  The full equations also lack growing modes for $k_z = 0$.  Fluid motions in this case are quite limited, especially given gas incompressibility.  Equations (\ref{v}, \ref{dv}) show that $w = \Delta w = 0$.\footnote{If $s = -1/(f_g \ts)$, then $\Delta w \ne 0$ is possible, but this damped mode is not of particular interest.}   The gas incompressibility condition, $k_x u_g = 0$, requires $u_g  = 0$ for a non-trivial mode.  Thus gas velocities are one dimensional, azimuthal.  We must allow two dimensional waves, and three dimensional motion, to find secular instability.

\subsection{Equal Mass}\label{eqmass}
When the mass densities of solids and gas are equal, so that $f_p = f_g = 1/2$, the constant term in (\ref{cubic}) vanishes.   In this case we have a static mode, $\omega = 0$, and the quadratic roots:
\begin{equation}
{-\imath \omega \over \Omega} = { \tau_s K_x^2\over 4 K^2}\left[-1 + \imath {K^2 \over K_x} \pm \sqrt{\left(1 - \imath {K^2 \over K_x}\right)^2 - \left({4K_z K\over \tau_s K_x^2}\right)^2} \right].
\end{equation}
The growth rate $s  \leq 0$ for for any (real) choice of parameters.\footnote{This follows trivially from the fact that $\Re(\sqrt{(1 + \imath a)^2 - b^2}) \leq 1$ for all real values of $a$ and $b$.}  Notice that as $\tau_s \rightarrow 0$ the modes approach the modified epicyclic frequency, $\omega/\Omega \rightarrow \pm K_z/K$ (this is the frequency of inertial oscillations in a single incompressible fluid).  Numerical solutions of the full equations give suppressed, but actually non-zero, growth when densities are nearly identical.  Also small-wavelengths ($K \gtrsim 1/\tau_s$) behavior, in which growth rates increase near equal densities, see figure \ref{KzInf}, is not captured in the terminal velocity approximation, as terms giving small scale accelerations have been dropped.

\subsection{Series Solutions}\label{loworder}
Aside from the special cases above, it is more enlightening to consider solutions to the approximate dispersion relation, equation (\ref{cubic}), as
a series expansion in $\tau_s \ll 1$:
\begin{equation}
\omega = \omega_0 + \omega_1\tau_s + \omega_2\tau_s^2 + ... \, .
\end{equation}
(Series solutions of the full sixth order dispersion relation have been done, but are not presented here.)  Two of the three modes described by equation (\ref{cubic}) are epicycles (inertial oscillations) to leading order, with $\omega_0/\Omega = \pm K_z/K$.  The first order correction, $\omega_1/\Omega = -f_gf_p K_x-\imath f_p (K_x/K)^2/2$ shows that these modes are damped to lowest order.    With the full set of equations, epicycles can grow for $K_z \gg K_x$, but growth rates of the secular mode are always faster.

The third, secular root is oscillatory to leading order:
\begin{equation}\label{cubicphase}
{\omega_1 \over \Omega} \approx 2 f_g (f_p - f_g) K_x  \, . 
\end{equation}
Thus the leading order wave speed is $\omega_1 \tau_s/k_x = \Usum$,
the sum of the equilibrium drift speeds of gas and solids, see equation (\ref{Usum}).  This agrees with the wave speed of growing modes for small $K_x$, before higher order corrections become significant.  Since $\omega_2 = 0$ for this mode,  $\omega_3$ gives the leading order growth rate:
\begin{equation}\label{cubgrow}
s_3 \equiv -\imath \Im(\omega_3) = 4 f_p f_g^2(f_p - f_g)^2  {K_x^4 \over K_z^2} \Omega \, .
\end{equation}
This rate is always positive, but nominally third order in $\tau_s \ll 1$.  However growth rates are larger for $K_x >> K_z$.  The growth is maximized at $K_z \approx \tau_s K_x^2$ as a higher order expansion (and e.g.\ figure \ref{kxkz}, \emph{top}) shows.  This asymmetry in wavenumbers makes the growth rate first order in $\tau_s$.

These low order expansions confirm some basic results about growth rates and wave speeds.  Most importantly, we demonstrate that the secular mode is responsible for fastest growth, not the pair of epicycles.

\section{Applications}\label{sec:app}

\subsection{Particle Concentration}\label{conc}
\begin{figure}[htb]
\begin{center}
\centerline {
\includegraphics[width=4in]
{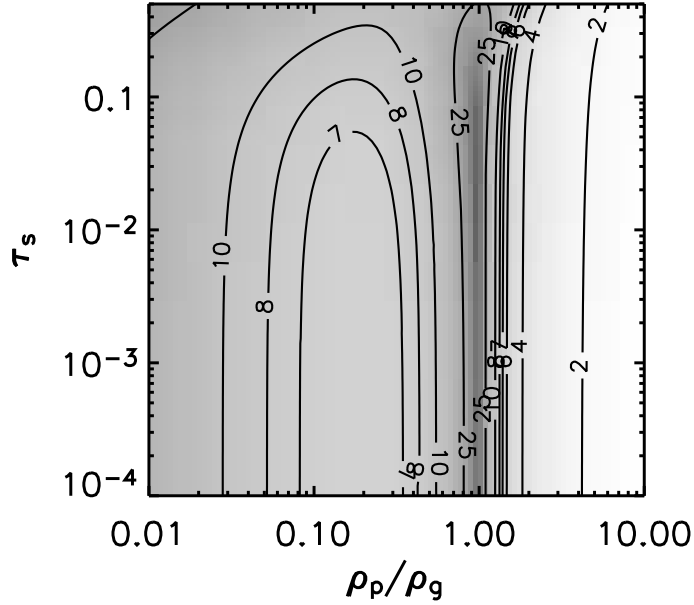}
}
\caption{Contours  of $A_\delta/A_h$, the ratio of particle density perturbations to radial pressure gradient perturbations,  in $\tau_s - \rho_p/\rho_g$ space for the growing modes of figures \ref{maxgrowth}.  Density perturbations are sizeable for $\rho_p/\rho_g \lesssim 1$.  The largest relative density perturbations, around $\rho_p/\rho_g \approx 1$, correspond to slowly growing modes.}
\label{density}
\end{center}
\end{figure}
Two fluid instabilities like ours could aid planetesimal formation by generating particle density perturbations.  With the aid of self-gravity these perturbations could eventually collapse to solid densities.  Perturbation amplitudes are arbitrary in a linear analysis, so inferences about non-linear development are speculative.  To estimate the relevance of density perturbations, we compare the perturbation amplitudes of particle density, $A_\delta \equiv |\delta|/f_p$  and radial pressure gradients, $A_h \equiv |k_x h/g_e|$. Figure \ref{density} shows that $A_\delta > A_h$, suggesting that density perturbations are significant.  By comparison with figure \ref{maxgrowth}, we see that the regions of largest growth rates do not coincide, and are somewhat anti-correlated actually, with the largest density perturbations.

We briefly justify using radial pressure gradients as the scale to compare the density perturbations.  Vertical gradients of pressure perturbations are smaller (since $k_z < k_x$) and more importantly have no background value in our unstratified model.  Perturbation velocities have six components and can be compared to several different background speeds, including  drift, Keplerian shear, and $\eta v_K$, the amplitude of  pressure supported sub-Keplerian rotation.  However the amplitudes $|{\bf v}|/\eta v_K$ and $|\Delta{\bf v}|/\Delta U$ are similar to $A_h$, i.e.\ somewhat smaller than $A_\delta$.

The mass of solids in an unstable mode varies considerably over the wide range of possible wavenumbers.   Let us conservatively take a small-scale mode with $K_x \sim K_z \sim 100$, in which the solid mass is
\begin{equation}
M_k =  {\Sigma_p \over H_p} {2 \pi \over k_x}{2 \pi \over k_z}{2 \pi \over k_y} \sim {10^{20} \over k_y \eta r}\, {\rm g},
\end{equation}
where the particle surface density $\Sigma_p \sim 10\, {\rm g/cm}^3$, $\eta \sim 10^{-3}$, and the particle scale height, $H_p \approx \eta r$, is the value for stirring by Kelvin Helmholtz instabilities (refs).  This is thin enough so that $\rho_p \gtrsim 0.1 \rho_g$.  Since our modes are initially axisymmetric, $k_y \eta r < 1$ is possible, but even if azimuthal breakup occurs on scales comparable to the radial wavelength, $k_y \eta r \sim K_x \sim 100$, $M_k$ is comparable in mass to a 100 km planetesimal.   Thus while non-linear development is unclear, the density perturbations induced by streaming instabilities contain more than enough mass to make healthy planetesimals.

\begin{figure}[htb]
\begin{center}
\centerline {
\includegraphics[width=4in]
{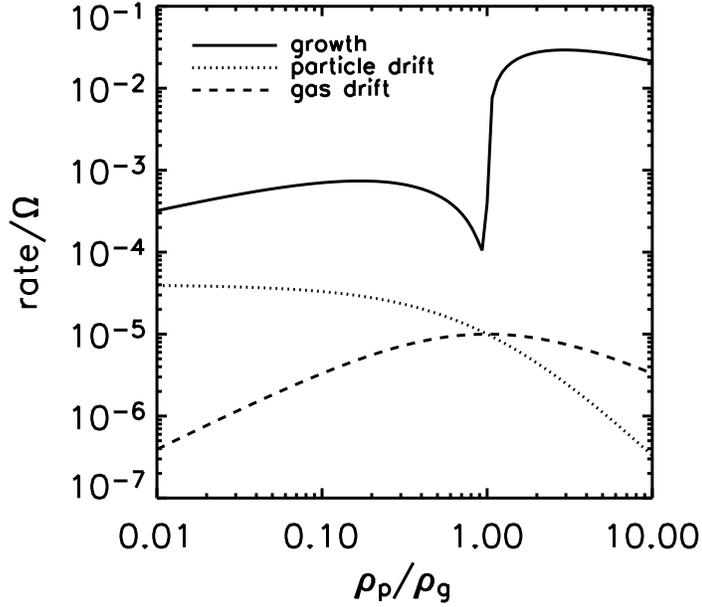}
}
\caption{Growth rates and equilibrium drift rates of solids and gas versus the density ratio, $\rho_p/\rho_g$ for $\eta = 2 \times 10^{-3}, \tau_s = 0.01$. 
The growth rates are faster than the drift rates for the given values.  This is true more generally as well, see text.}
\label{drift_vs_growth}
\end{center}
\end{figure}

\subsection{Growth vs. Drift Rates}\label{sec:gvd}
Our local treatment of the instability is valid only if the growth is faster than global disk evolution, including changes in surface density or temperature.  Single fluid accretion disk models evolve on timescales $>10^5 - 10^6$ years, while passive disks evolve more slowly.   Global redistribution of solids, at the equilibrium radial drift speed $U_p$, changes particle surface densities (YS, YC).  Gas densities are less subject to change because drift speeds are smaller (when $\rho_p < \rho_g$) and because drift rates are  much smaller for the majority of the gas mass, which lies outside the particle-dense midplane.

To justify the local treatment of the instability, we compare growth rates to equilibrium radial drift, which leads to a global redistribution of solids (YS, YC).
Figure \ref{drift_vs_growth} shows that growth rates are at least an order of magnitude faster than the particle drift rate, $U_p/r$.  Gas drift rates, also slower than growth rates, are shown as well.  In a stratified disk, the gas drift rate will decrease with height as the particle concentration drops, while the particle drift rate asymptotes to a constant value.
For all reasonable values of stopping time and pressure support, $\eta$, growth should still dominate.  Both the growth and drift rates are linearly proportional to stopping time for $\tau_s \ll 1$.  A hotter disk, i.e.\ larger $\eta \sim c_g^2/v_K^2 \propto T$, has a faster drift rate, while growth rates are unaffected by $\eta$.\footnote{The fastest growing wavelength is shorter in a hot disk, with $k \propto \eta$.}  As long as disks are not hotter than standard models by  more than an order of magnitude  (a safe assumption according to observations), we conclude that growth rates are robustly faster than drift rates.  This enhances our confidence that the instability can be astrophysically relevant.



\section{Other Work}\label{comp}
There have been several previous studies of the linear stability of
gas-and-dust mixtures in the context of the protosolar nebula.  These
works include physical effects that we have neglected, such as
self-gravity and vertical stratification.  But because of various
restrictions on the types of perturbations considered, none found the
modes described in this paper.  \cite{Coradini_Magni_Federico81} and
\cite{Noh_Vishniac_Cochran91} wrote down two-fluid equations including
self-gravity but permitted only horizontal motions.  \cite{Sekiya83}
allowed vertical motions but worked in the tightly-coupled limit where
the velocity difference between the two fluids is negligible.  All of these
authors found gravitational instabilities at sufficiently high densities.
\cite{is03} and \cite{Garaud_Lin04} examined the
stability of the vertical shear between the settling dust layer and
the overlying gas but also neglected the slippage between the two
fluids.

Like the present paper, \cite{Goodman_Pindor00} found an instability
driven by drag rather than self-gravity, but theirs was not a complete
two-fluid analysis.  Following \cite{gw73}, GP treated the
dust as a monolithic though dilute layer, the drag being exerted at
its top and bottom surfaces by boundary-layer turbulence driven by the
difference in orbital velocity between the dust-laden and dust-free
components.  The dynamics of the gas was not treated explicitly, its
effects on the dust layer being parametrized in terms of the orbital
velocity difference.  GP did however emphasize that they expected the
existence (though not the growth rate) of drag instabilities to be
independent of many physical details provided that the drag is a
collective effect: in other words, that the drag on a dust particle
depends upon neighboring particles and is not linearly
proportional to dust mass.  In GP's case, the collective property
derives from the assumption that the drag depends on conditions at the
surface of the dust layer only, so that the average drag per particle
is inversely proportional to the column density of the dust layer.
We, however, have resolved the vertical dimension explicitly, so that
the drag on a small dust particle depends on its motion relative to
the local gas only.  But because of the backreaction of the dust on
the gas, the drag is collective: as the local ratio $\rho_{\rm
p}/\rho_{\rm g}$ increases, the relative velocity and the drag per
particle decrease.  Apparently, this is enough to support an instability
despite the many simplifications assumed for our background state.

\section{Conclusion}\label{disc}
We describe a two-fluid streaming instability relevant to protoplanetary disks of particles and gas.  Unstable modes are powered by the relative drift between the two components, a universal consequence of radial pressure gradients.  Rotation, which is Keplerian in our model, is another necessary ingredient.  Growth occurs despite the fact that the two components interact only via dissipative drag forces.  The robust instability has growth rates slower than dynamical times but faster than drift times.  The fluid motions generate particle density enhancements, even in the absence of self-gravity, which could trigger planetesimal formation.  The physics of our disk model was simplified considerably (see \S\ref{basic}).  Numerical studies that include vertical stratification, a dispersion of particle sizes,  and non-linear effects could elucidate the role of such instabilities in protoplanetary disk evolution.

\acknowledgments

We are grateful to Andrew Cumming, Marc Kuchner, Doug Lin, Gordon Ogilvie, and Scott Tremaine for helpful comments, many of which were made during the successful KITP workshop on Planet Formation.  This material is based upon work supported by the National Aeronautics and Space Administration under Grant NAG5-11664 issued through the office of space science.  This research was supported in part by the National Science Foundation under Grant No. PHY99-07949.

\appendix

\section{Simplified Equations of Motion}
Derivation of the cubic dispersion relation, equation (\ref{cubic}), uses the terminal velocity approximation, equation (\ref{dvsimp}), with equations (\ref{v}, \ref{cont}, and \ref{incomp}) and neglecting the ${\bf F}'$ and ${\bf G}'$ terms.  Directly solving for the characteristic equation of this set introduces terms, including a quartic in $\omega$, which must be dropped for consistency with the $\tau_s \ll 1$ approximation.  Alternatively, elimination of pressure and relative motion variables gives the following three equation set:
\begin{eqnarray}
(\omega - f_g k_x \Delta U)\delta &=& 2 f_p k_x \tau_s v \label{delvcubic}\\
-\imath \omega \Gamma - 2\Omega k_z v &=& k_z \Omega \Delta U \delta/\tau_s \\
\left(-\imath \omega + {k_x^2 \over k^2} f_p \tau_s \Omega \right)v + {k_z \over 2 k^2} \Omega \Gamma &=& -{k_x^2 \over 2 k^2}f_g \Omega \Delta U \delta
\end{eqnarray}
where $\Gamma \equiv k_z u - k_x w$, is proportional to azimuthal vorticity (modulo a phase).  The dispersion relation follows directly.  Equation (\ref{delvcubic}) gives a simple relation between density and azimuthal velocity perturbations, which are nearly in phase since the growth rate is usually small compared to $\omega_\Re - f_g k_x \Delta U$.
\bibliography{refs}
\end{document}